\documentclass[sigconf, nonacm,natbib=false]{acmart}
\usepackage{subfigure}
\usepackage{acronym}
\usepackage{comment}
\usepackage{makecell}

\copyrightyear{2025}
\acmYear{2025}
\setcopyright{acmlicensed}\acmConference[NANOCOM '25]{The 12th Annual ACM International Conference on Nanoscale Computing and Communication}{October 23--25, 2025}{Chengdu, China}
\acmBooktitle{The 12th Annual ACM International Conference on Nanoscale Computing and Communication (NANOCOM '25), October 23--25, 2025, Chengdu, China}
\acmDOI{10.1145/3760544.3764129}
\acmISBN{979-8-4007-2166-3/2025/10}

\usepackage[style=ieee,backend=bibtex,,maxbibnames=10]{biblatex}
\bibliography{biblio}

\begin{document}

\title{Set Transformer Architectures and Synthetic Data Generation for Flow-Guided Nanoscale Localization}


\author{Mika Leo Hube}
\affiliation{%
  \institution{NaNoNetworking Center in Catalunya, Polytechnic University of Catalonia, Spain}
  \country{}}
\email{mika.leo.hube@estudiantat.upc.edu}

\author{Filip Lemic\footnotemark\authornote{Corresponding author.}\footnotemark\authornote{F. Lemic is also with Faculty of Electrical Engineering and Computing, University of Zagreb, Croatia.}}
\affiliation{%
  \institution{AI-driven Systems Lab, i2Cat Foundation, Spain}
  \country{}}
\email{filip.lemic@i2cat.net}

\author{Ethungshan Shitiri}
\affiliation{%
  \institution{NaNoNetworking Center in Catalunya, Polytechnic University of Catalonia, Spain}
  \country{}}
\email{ethungshan.shitiri@upc.edu}

\author{Gerard Calvo Bartra}
\affiliation{%
  \institution{AI-driven Systems Lab, i2Cat Foundation, Spain}
  \country{}}
\email{gerard.calvo@i2cat.net}

\author{Sergi Abadal}
\affiliation{%
  \institution{NaNoNetworking Center in Catalunya, Polytechnic University of Catalonia, Spain}
  \country{}}
\email{abadal@ac.upc.edu}

\author{Xavier Costa Pérez\footnotemark\authornote{X. Costa is also with NEC Laboratories Europe GmbH, Germany and ICREA, Spain.}}
\affiliation{%
  \institution{AI-driven Systems Lab, i2Cat Foundation, Spain}
  \country{}}
\email{xavier.costa@i2cat.net}

\renewcommand{\authors}{M. L. Hube, F. Lemic, E. Shitiri, G. Calvo Bartra, S. Abadal, X. Costa Pérez}

\renewcommand{\shortauthors}{Hube et al.}

\begin{abstract}
\ac{FGL} enables the identification of spatial regions within the human body that contain an event of diagnostic interest. 
\ac{FGL} does that by leveraging the passive movement of energy-constrained nanodevices circulating through the bloodstream. 
Existing \ac{FGL} solutions rely on graph models with fixed topologies or handcrafted features, which limit their adaptability to anatomical variability and hinder scalability. 
In this work, we explore the use of Set Transformer architectures to address these limitations. 
Our formulation treats nanodevices' circulation time reports as unordered sets, enabling permutation-invariant, variable-length input processing without relying on spatial priors.
To improve robustness under data scarcity and class imbalance, we integrate synthetic data generation via deep generative models, including \acs{CGAN}, \acs{WGAN}, \acs{WGAN-GP}, and \acs{CVAE}. 
These models are trained to replicate realistic circulation time distributions conditioned on vascular region labels, and are used to augment the training data. 
Our results show that the Set Transformer achieves comparable classification accuracy compared to \ac{GNN} baselines, while simultaneously providing by-design improved generalization to anatomical variability. 
The findings highlight the potential of permutation-invariant models and synthetic augmentation for robust and scalable nanoscale localization.
\end{abstract}

\ccsdesc[500]{Networks~Network architectures}
\ccsdesc[500]{Networks~Network simulations}
\ccsdesc[500]{Networks~Location based services}
\ccsdesc[300]{Human-centered computing}

%
\keywords{Cardiovascular precision medicine, in-body nanoscale communication, flow-guided localization, generative artificial intelligence.}

\maketitle



\acrodef{ML}{Machine Learning}
\acrodef{THz}{Terahertz}
\acrodef{GNN}{Graph Neural Networks}
\acrodef{ZnO}{Zinc Oxide}
\acrodef{IMU}{Inertial Measurement Unit}
\acrodef{RF}{Radio Frequency}
\acrodef{SINR}{Signal to Interference and Noise Ratio}
\acrodef{NN}{Neural Network}
\acrodef{ctDNA}{circulating tumor DNA}
\acrodef{FGL}{Flow-guided Localization}
\acrodef{HGT}{Heterogeneous Graph Transformer}
\acrodef{ISAB}{Induced Set Attention Block}
\acrodef{PMA}{Pooling by Multihead Attention}
\acrodef{SAB}{Set Attention Block}
\acrodef{CGAN}{Conditional Generative Adversarial Network}
\acrodef{WGAN}{Wasserstein Generative Adversarial Network}
\acrodef{WGAN-GP}{Wasserstein GAN with Gradient Penalty}
\acrodefplural{WGAN-GP}[WGAN-GPs]{Wasserstein GANs with Gradient Penalty}
\acrodef{CVAE}{Conditional Variational Autoencoder}
\acrodef{GMM}{Gaussian Mixture Model} 

\section{Introduction}

Nanodevices integrating sensing and communication are considered as an enabler for advanced medical diagnostics. 
Among the most promising applications is \acf{FGL}, which leverages the passive mobility of such devices within the human bloodstream to associate sensed events with their approximate spatial origin~\cite{gomez2022nanosensor,simonjan2021body,bartra2024graph}. 
This enables non-invasive and cost-efficient localization of disease markers and anomalies, offering clear benefits in terms of early diagnostics and reduced procedural risks~\cite{chen2020nanotechnology,diehl2005detection}. 
Recent studies have shown that even coarse localization at the region level can deliver actionable medical insights, for instance, in cancer detection or circulatory disease screening~\cite{song2023vitro,swietach2014chemistry}.

Despite its potential, \ac{FGL} poses multiple challenges. 
Nanodevices are extremely resource-constrained and can only report limited telemetry to external anchors via short-range \ac{THz}-based backscattering~\cite{lemic2021survey}. 
As such, localization must be inferred externally from sparse, noisy observations typically involving circulation times and binary event bits, collected at a heart-adjacent anchor.
Previous works have addressed this by training ML models to classify the region traversed by each nanodevice during circulation~\cite{gomez2022nanosensor,simonjan2021body,bartra2024graph,lemic2024insights}. 
However, these approaches often rely on fixed-length feature vectors or handcrafted graph structures, limiting their generalization and scalability.
Moreover, obtaining large-scale, diverse, and well-labeled datasets for \ac{FGL} is inherently difficult due to the complexity and variability of in-body conditions. 
This leads to challenges such as class imbalance and data sparsity, which can degrade model performance and limit generalization.

\begin{figure*}
\centering
\includegraphics[width=\linewidth]{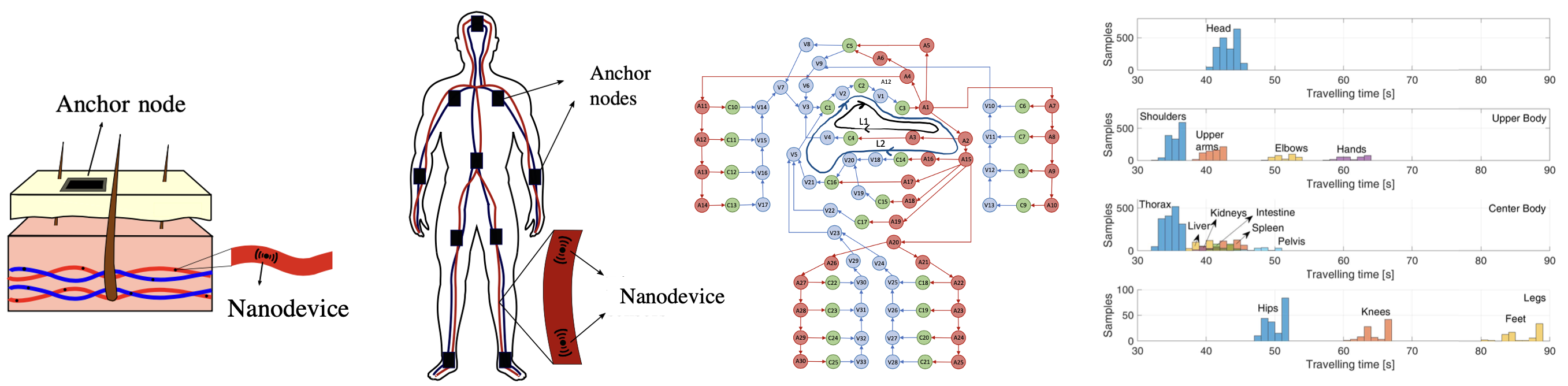}
\vspace{-3mm}
\caption{Main components of the considered flow-guided localization model: (1) Nanodevices passively circulate through a simulated vascular network and report their circulation times to a heart-adjacent anchor via THz backscattering; (2) the framework supports extension to multi-anchor scenarios to enhance spatial resolution and disambiguate symmetric regions; (3) the mobility of nanodevices is modeled as a Markov process, where each region is associated with a transition probability, characteristic flow speed, and traversal distance; (4) the system collects iteration time reports (i.e., time-of-return) for each region, which serve as inputs to the localization model.}
\label{fig:vision}
\vspace{-2mm}
\end{figure*}   

In this work, we explore a more flexible formulation of the localization problem by modeling each nanodevice's reports as unordered sets of variable length.
The considered \ac{FGL} model is depicted in Figure~\ref{fig:vision}.
To this end, we employ the Set Transformer architecture~\cite{lee2019set}, which supports permutation invariance and set-level reasoning through self-attention mechanisms. 
This eliminates the need for anatomical priors or graph construction heuristics. 

To address data scarcity and class imbalance, we explore the use of synthetic data generation techniques that can simulate realistic nanodevice behavior under diverse anatomical scenarios, thereby enhancing the robustness of learning models trained on limited or skewed datasets.
We use deep generative models, specifically, \acp{CGAN}, \acp{WGAN}, \acp{WGAN-GP}, and \acp{CVAE}, conditioned on region labels~\cite{mirza2014conditional,arjovsky2017wasserstein,gulrajani2017improved,goodfellow2020generative}.

Our results show that Set Transformer models can match or outperform existing baselines based on \acp{GNN}, while offering improved flexibility and reduced reliance on structured input representations. 
Moreover, augmenting training data with samples generated by conditional generative models enhances their generalization capabilities.
These findings motivate further exploration of set-based modeling and generative augmentation for robust and scalable nanoscale localization systems.

\begin{figure*}
\centering
\includegraphics[width=0.84\linewidth]{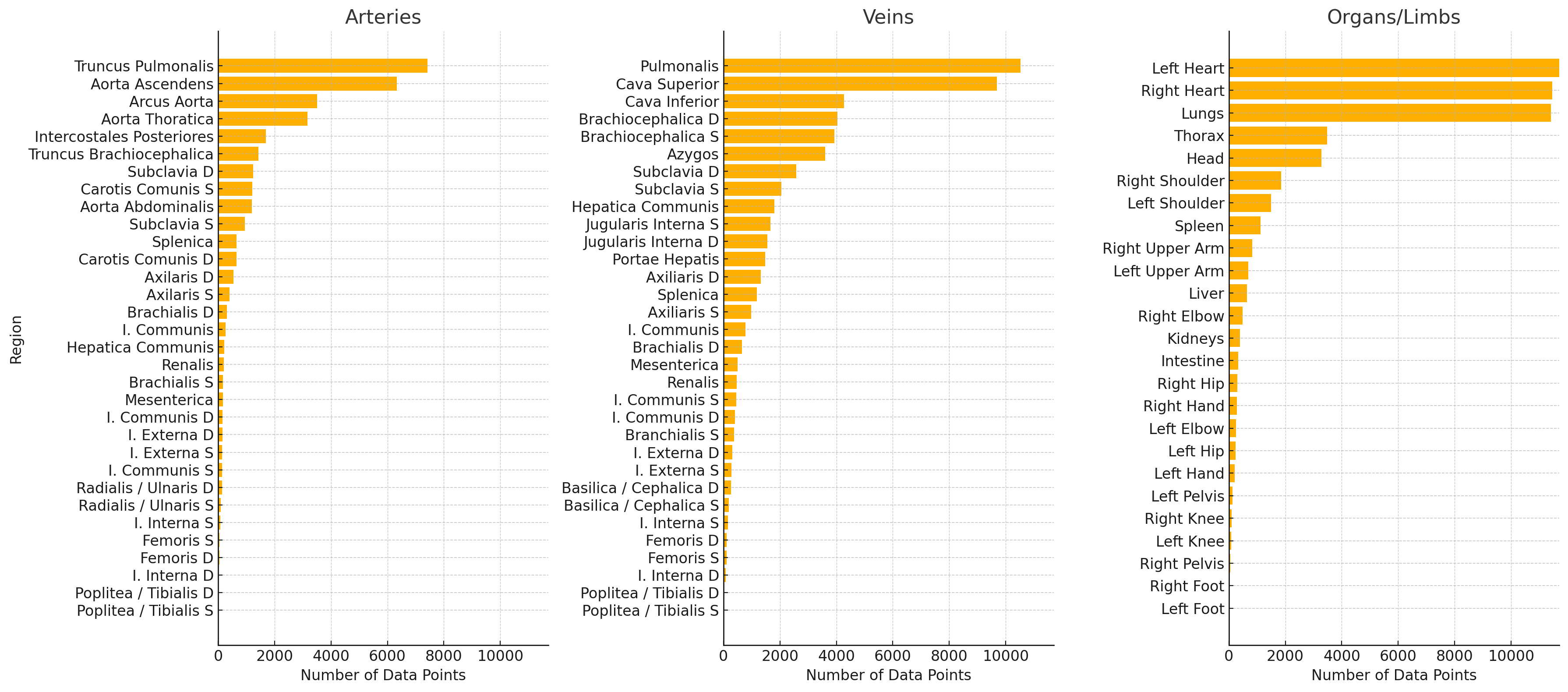}
\vspace{-2mm}
\caption{Region class imbalance}
\label{fig:25_regions}
\vspace{-2mm}
\end{figure*}   




The structure of this paper is as follows. 
Section~\ref{sec:related_works} provides an overview of relevant \ac{FGL} models and driving use-cases. 
Section~\ref{sec:design} presents the system design, including the considered Set Transformer model, generative methods, and augmentation strategies. Evaluation results from Section~\ref{sec:results} compare the performance of the proposed design against \ac{GNN} baselines, followed by conclusions and future research directions.


\section{Background and Related Works}
\label{sec:related_works}

\subsection{Use Cases}

\ac{FGL} has been proposed as a foundation for in-vivo detection and localization of medical conditions and events. 
\ac{FGL} is expected to be beneficial for cancer detection, in particular for identifying elevated concentrations of \ac{ctDNA} near tumor sites, exploiting its short half-life and localized presence~\cite{chen2020nanotechnology,diehl2005detection}. 
Cardiovascular diseases such as deep vein thrombosis and pulmonary embolism have also been highlighted as suitable targets, with nanodevices envisioned to detect localized blood pressure anomalies and biomarker levels including D-Dimer, PAI-1, and soluble fibrin~\cite{schaffer201017}. 
For acute myocardial infarction, elevated troponin levels localized to the heart can serve as reliable indicators, enabling timely detection through \ac{FGL}-based sensing~\cite{thygesen2018fourth}.

Beyond direct disease detection, \ac{FGL} has been suggested as a general-purpose diagnostic tool for continuous monitoring of physiological parameters such as body temperature, blood pressure, and pH~\cite{chen2019thermometry,lahiri2012medical,song2023vitro,swietach2014chemistry}. 
By associating sensed data with localized vascular regions via circulation-based timing, nanodevices may support early diagnosis of inflammation, ischemia, or cancer-related metabolic changes. 
Moreover, \ac{FGL} can be used to detect circulatory anomalies.
For instance, a lack of nanodevice returns to the heart-adjacent anchor may indicate vascular obstructions, as in cases of severe vasoconstriction. 
While these use cases demonstrate the broad potential of \ac{FGL}, they underscore the need for robust simulation environments and standardized performance evaluation to validate the reliability and accuracy of such approaches.

\subsection{Flow-guided Localization Models}

Notable \ac{FGL} models are~\cite{gomez2022nanosensor,simonjan2021body}, in which \ac{ML} models are employed to differentiate the regions through which each nanodevice passes during a single circulation through the bloodstream. 
The approach in~\cite{simonjan2021body} estimates the path of each nanodevice by using distance data collected from a conceptual nanoscale \ac{IMU}, though this strategy is constrained by the limited resources for data storage and processing onboard the nanodevice, and by the sensitivity of \ac{IMU} measurements to vortex flows in the bloodstream. 
In contrast, the method in~\cite{gomez2022nanosensor} tracks the circulation time between consecutive heart passages and transmits this timing information to an on-body anchor via short-range \ac{THz}-based backscattering, mitigating some of the hardware and energy limitations associated with \ac{IMU}-based tracking.

\acp{GNN} have emerged as a promising candidate for \ac{FGL} due to their capacity to model vascular systems as graph structures~\cite{bartra2024graph}. 
In this representation, region nodes (corresponding to anatomical areas) and anchor nodes (representing communication points) are connected through edges informed by vessel topology and flow dynamics. 
Feature vectors for these nodes are derived from circulation time statistics and event bit patterns collected by nanodevices. 
The use of \ac{HGT} layers enables the model to handle heterogeneous node types and capture message-passing dynamics between anchors and body regions. 

However, \ac{GNN}-based models face several limitations that restrict their generalization to real-world deployments. They require prior anatomical knowledge (e.g., vessel lengths, flow velocities) that may be unavailable or patient-specific in practice. Next, they compress variable-length circulation time data into summary statistics (e.g., via \acp{GMM}), potentially losing high-resolution information critical to accurate localization~\cite{bartra2024graph}. Third, the underlying graphs remain static and handcrafted, limiting adaptability to anatomical variability. 
Finally, distinguishing left/right body symmetry is inherently challenging due to similar circulation times, making localization ambiguous without additional anchors. 

These issues highlight the need for alternative architectures that natively support permutation-invariant, variable-length inputs, and that minimize reliance on hardcoded spatial priors, motivating the exploration of transformer-based models in \ac{FGL}.
Our work contributes to this evolving research landscape by introducing permutation-invariant architectures and synthetic data augmentation as promising directions for more robust, scalable, and generalizable nanoscale \ac{FGL} systems.

\section{Methodology}
\label{sec:design}

This section presents the methodology applied in our study, structured into three key components: i) Set Transformer model architecture employed for region-level classification, ii) generative models used for synthetic data generation, and iv) strategies adopted for data augmentation. All models are evaluated using a standardized simulation framework that mirrors the assumptions from~\cite{bartra2024graph}, with a heart-based anchor and variable nanodevice circulation behavior as described in Section~3.
Hyperparameter optimization was generally conducted using Optuna~\cite{akiba2019optuna}.

\subsection{Set Transformer Architecture}

Set Transformer~\cite{lee2019set} is employed to address the limitations of \acp{GNN}~\cite{bartra2024graph}, particularly their reliance on static graph structures and anatomical priors. 
Set Transformer architectures natively support variable-length, unordered input sets, which is an ideal fit for \ac{FGL}, where circulation times vary across simulations and their ordering is non-informative. 
The model incorporates permutation invariance and avoids hand-crafted summarization, relying instead on self-attention mechanisms to model set-level interactions.

Our implementation utilizes two \ac{ISAB} layers in the encoder and a \ac{PMA} layer for set pooling in the decoder. 
The final prediction is produced via a \ac{SAB} and a linear output layer. 
Training is conducted with the Adam optimizer and a cosine annealing learning rate scheduler. 
Hyperparameter tuning is performed using the Tree-structured Parzen Estimator~\cite{watanabe2023tree}, with the search orchestrated via Ray Tune~\cite{liaw2018tune}. 
The search space includes learning rate, number of attention heads, inducing points, hidden dimensions, dropout rate, and the use of AMSGrad. 
Each sample is normalized circulation time values derived from nanodevice outputs, and training is done per set due to the lack of batching support, which increases training time.

\subsection{Synthetic Data Generation}

To improve generalization, we explore deep generative models for synthesizing realistic circulation time distributions, conditioned on region labels. 
The candidates include \acp{CGAN}~\cite{mirza2014conditional}, \acp{WGAN}~\cite{arjovsky2017wasserstein}, \acp{WGAN-GP}~\cite{gulrajani2017improved}, and \acp{CVAE}~\cite{goodfellow2020generative}. 
These models are trained on real-valued scalar circulation time samples, with conditioning applied via label embeddings.

Evaluation is based on the average Wasserstein distance between real and generated circulation time distributions, computed independently per region. 
For CGANs, the generator and discriminator are implemented as fully connected networks with embedded label conditioning. 
CVAEs use a single hidden layer in both encoder and decoder, with latent variables sampled via the reparameterization trick. 
Hyperparameter optimization for CGAN and CVAE is conducted over learning rates, layer widths, latent dimensions, and embedding sizes.

\subsection{Data Augmentation}

To further explore the design space, we evaluate the impact of synthetic data on classification performance through dynamic data augmentation during training. 
The key idea is to enrich the sets with additional generated samples while preserving the distributional structure. 
Two augmentation strategies are considered: (i) \textit{fixed}, where half the original set size is appended with synthetic values, and (ii) \textit{random}, where the number of synthetic samples per set is uniformly sampled from 0 to the original set size. 
The SetTransformer directly ingests the augmented set, while the \ac{GNN} uses GMM-derived features recomputed from the new set.

This approach ensures consistency between architectures while allowing us to evaluate whether synthetic samples improve generalization, particularly under class imbalance or sparse sampling regimes. 
Samples are generated on-the-fly rather than precomputed to increase variability and robustness.


\section{Evaluation Results}
\label{sec:results}

\subsection{Evaluation Setup}

Our evaluation follows a standardized benchmarking protocol tailored to region-level event localization in the human bloodstream. 
All models are trained and tested using synthetic datasets generated by a simulator that mirrors the physiological and mechanical properties of the vascular system~\cite{lopez2024toward}. 
The utilized simulator addresses key in-body challenges associated with flow-guided localization, including limited \ac{THz} communication range, nanodevice energy constraints due to energy harvesting, and high mobility in dynamic blood flow environments~\cite{lemic2022toward}.
It supports event detection modeling, beacon-based reporting, and intermittent nanodevice behavior based on realistic energy harvesting cycles. Importantly, the simulator implements a workflow for objective benchmarking that standardizes evaluation metrics and experimental scenarios. 
This allows for reproducible, scenario-independent comparisons across \ac{FGL} solutions and aims to prevent early-stage inconsistencies in evaluating in-body localization methods.

\begin{table}[!t]
\centering
\caption{Simulation parameters}
\vspace{-2mm}
\small
\begin{tabular}{l l }
\hline
\textbf{Parameter} & \textbf{Value} \\ \hline
Anchor beaconing interval & 100 ms \\ 
Nanodevice sampling rate & 3 samples/s \\
Simulation duration & 1200 s \\ 
Event detection distance & 1 cm \\ 
Blood flow speed (aorta) & 20 cm/s \\ 
Blood flow speed (arteries) & 10 cm/s \\ 
Blood flow speed (veins) & 2 - 4 cm/s \\ 
Transition speed (organs/limbs/head) & 1 cm/s \\
Generator voltage \( V_g \) [V] & 0.42 \\ 
Energy consumed in pulse reception [pJ] & 0.0 \\
Energy consumed in pulse transmission [pJ] & 1.0 \\
Maximum energy storage capacity [pJ] & 800 \\ 
Turn ON/OFF thresholds [pJ] & 10/0 \\
Harvesting cycle duration [ms] & 20 \\
Harvested charge per cycle [pC] & 6 \\ 
Transmit power \( P_{TX} \) [dBm] & -20 \\ 
Operational bandwidth [GHz] & 10 \\ 
Receiver sensitivity [dBm] & -110 \\ 
Operational frequency [THz] & 1 \\ \hline
\end{tabular}
\label{tab:baseline_parameters}
\vspace{-5mm}
\end{table}

The simulator provides a simplified circulatory topology~\cite{geyer2018bloodvoyagers} composed of 94 interconnected vessels and organs. 
Each nanodevice passively circulates through this system and reports telemetry data to a static anchor located near the heart, which is assumed to be continuously powered and capable of receiving backscattered responses.
The task is formulated as a 94-class classification problem, where each class corresponds to an anatomical region (e.g., organs, veins, and arteries).

A nanodevice is considered to have detected an event if it is active and within a fixed distance threshold of the event location at the time of sampling. Training data is generated by placing one event per region at the region’s centroid and simulating nanodevice behavior for 1200 seconds. 
Test events are randomly placed within each region to simulate intra-class spatial variability. 
Evaluation metrics include region classification accuracy, point accuracy (defined as the Euclidean distance between estimated and true location coordinates), and training time.
The simulation parameters used in our study are based on~\cite{jornet2012joint} and summarized in Table~\ref{tab:baseline_parameters}.

As a baseline, we consider a graph-based localization model based on the architecture proposed in~\cite{bartra2024graph}. The vascular system is modeled as a directed graph composed of region and anchor nodes connected based on blood flow topology. Circulation time values are compressed via \acp{GMM}, and extracted statistical features serve as node attributes. A \ac{HGT} is trained on this representation to classify the region associated with a detected event.

We used a 90/10\% train-validation split for all model experiments. All experiments were conducted on a machine equipped with an Intel(R) Xeon(R) Gold 6326 @ 2.90GHz (10 cores) CPU, an NVIDIA RTX A5000 GPU with 24GB VRAM, and 24GB of system RAM.

\subsection{Hyperparameter Tuning}

\begin{figure}[!t]
\centering
\includegraphics[width=0.92\linewidth]{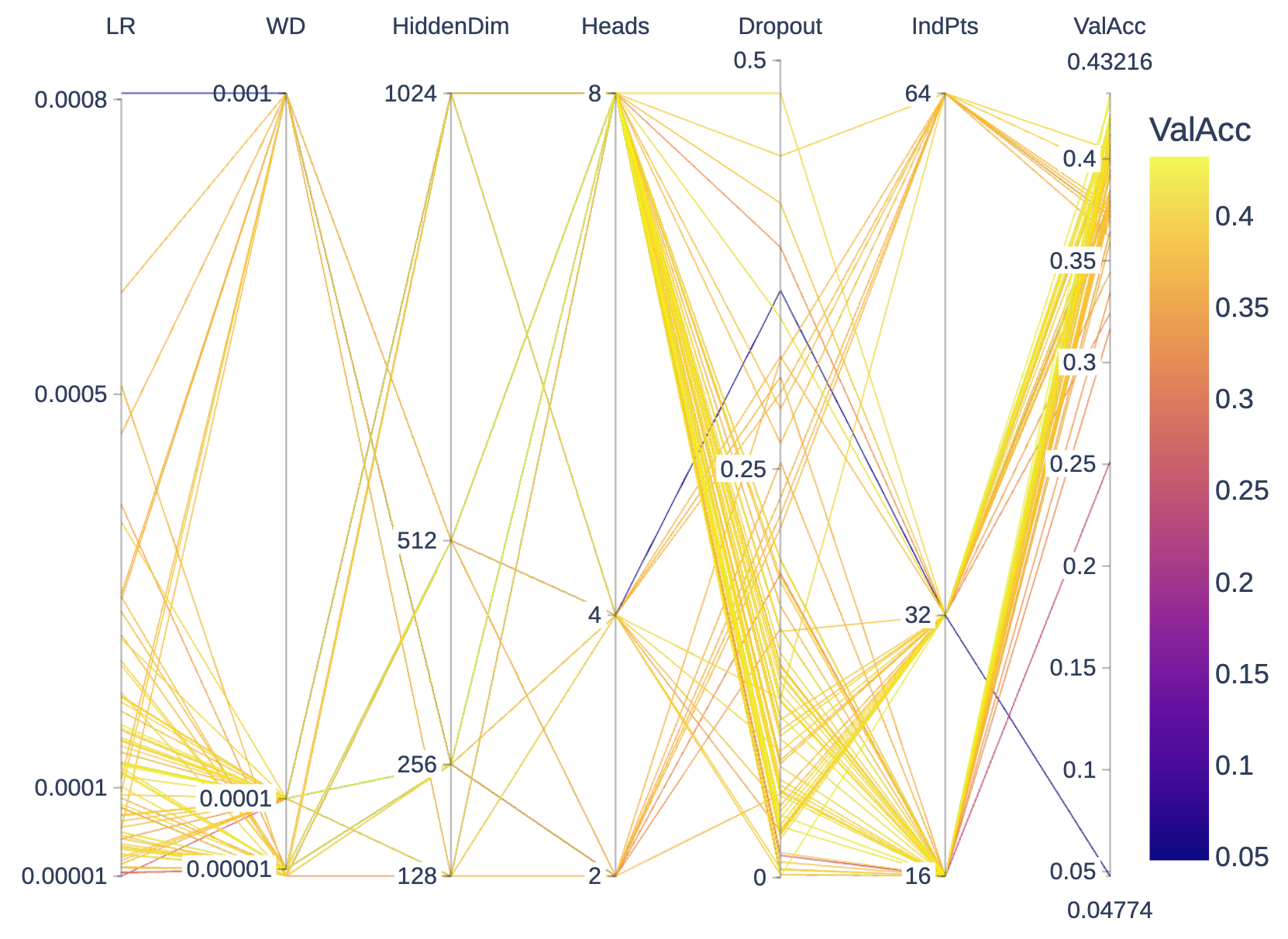}
\vspace{-2mm}
\caption{Set Transformer hyperparameter tuning}
\label{fig:hyperparameter}
\end{figure}

Each model underwent dedicated hyperparameter tuning to ensure fair comparison and optimal performance, as exemplified for the Set Transformer in Table~\ref{tab:search_space} and Figure~\ref{fig:hyperparameter}. 
The hyperparameter search space for the other models can be found in~\cite{hube2025exploring}. 
The GNN baseline models the circulatory system as a heterogeneous directed graph and uses GMM-derived statistical features for node attributes. We tuned the number of hidden layers (1–3), hidden layer width (64–512), learning rate ($10^{-5}$–$10^{-3}$), dropout (0–0.4), and number of GMM components (2–5). The model was trained using negative log-likelihood loss and Adam optimizer. The best configuration was selected based on validation set's region classification accuracy.

For the Set Transformer, we explored configurations spanning the number of inducing points (16–64) and attention heads (2–8), hidden dimension size (128–1024), dropout rate (0–0.5), and a binary indicator on the usage of AMSGrad. 
Given the lack of native batching for variable-length sets, each training sample was processed independently, resulting in lower utilization of computing resources. The final configuration achieved competitive region classification accuracy while reducing reliance on handcrafted features and anatomical priors.

For the CGAN, both the generator and discriminator were modeled as fully connected networks with label-conditioned inputs. Tuned parameters included hidden layer width (32–256), number of layers (1–3), learning rate ($10^{-5}$–$10^{-3}$), and embedding dimensions (4–16). WGAN and WGAN-GP shared the architecture but differed in loss function and training stability strategies. For WGAN-GP, the gradient penalty coefficient was also tuned. CVAE models included a latent dimension (2–16), hidden layer width (32–256), and conditioning strategy through label embeddings. All models were evaluated using the average Wasserstein distance between real and synthetic circulation time distributions across regions.

\begin{table}[!t]
\centering
\caption{Set Transformer hyperparameter search space}
\vspace{-2mm}
\label{tab:search_space}
\small
\begin{tabular}{@{}lc@{}}
\toprule
Parameter           & Search Space  \\
\midrule
Learning rate (LR)       & $[1e^{-5},\,1e^{-3}]$                         \\
Weight decay (WD)        & $\{1e^{-6},\,1e^{-5},\,1e^{-4},\,1e^{-3}\}$   \\
AMSGrad                  & \{True, False\}                               \\
Hidden dimensions        & \{128, 256, 512, 1024\}                       \\
Attention heads          & \{2, 4, 8\}                                   \\
Inducing points          & \{16, 32, 64\}                                \\
Dropout probability      & $[0,\,0.5]$                                   \\
Epochs                   & \{50, 75\}                                    \\
\bottomrule
\end{tabular}
\end{table}

The tuned models were subsequently used in the data augmentation pipeline, with synthetic samples generated during training This ensured variability and reduced the risk of overfitting to fixed synthetic patterns. Hyperparameter tuning played a critical role in stabilizing training, especially for adversarial models.

\subsection{Evaluation Results}

\begin{figure}[!t]
\vspace{-2mm}
    \centering
    \subfigure[CGAN Model]{\includegraphics[width=0.35\textwidth]{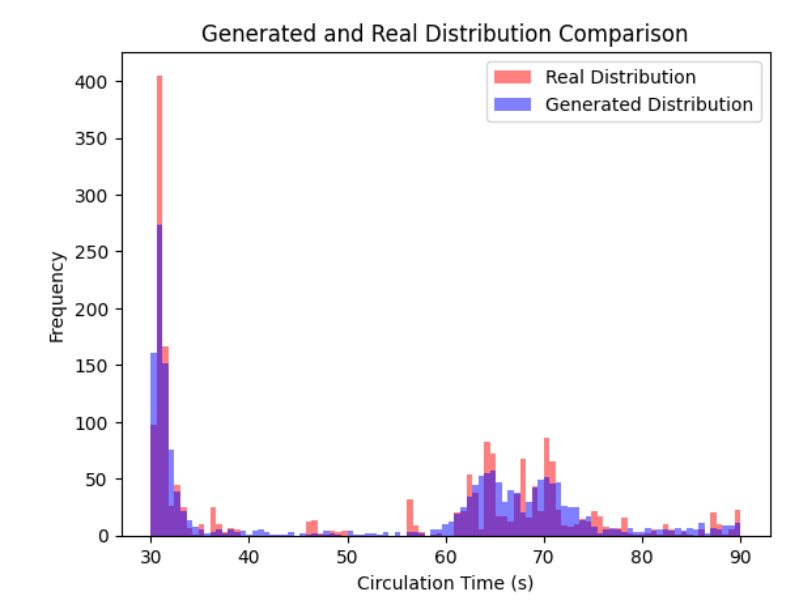}\vspace{-2mm}} 
    \subfigure[CVAE Model]{\includegraphics[width=0.35\textwidth]{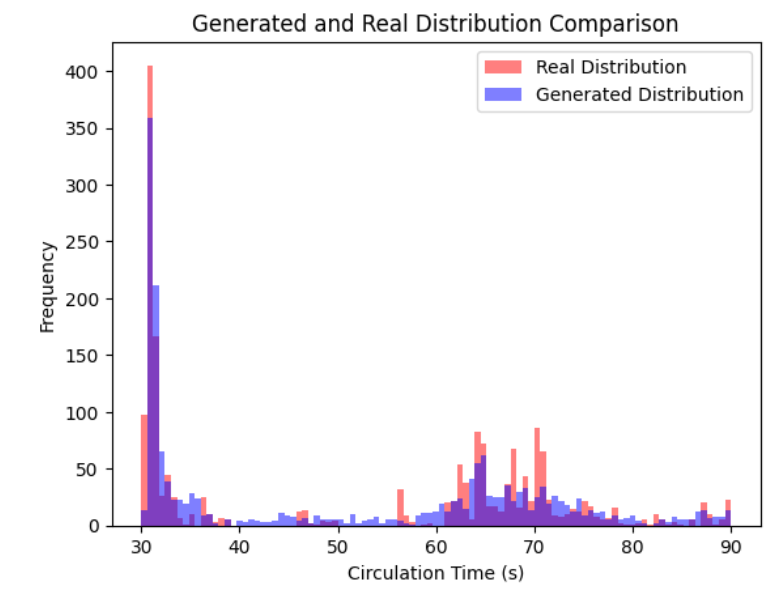}} 
    \vspace{-3mm}
    \caption{Generated and original distributions for the right shoulder region using optimized models}
    \label{fig:generated_distributions}
    \vspace{-5mm}
\end{figure}

Among the generative models, CGAN and CVAE produced the most realistic distributions validated via average Wasserstein distance. Figure~\ref{fig:generated_distributions} demonstrates that these models are capable of generating synthetic data with comparable distributions to real data.

As shown in Table~\ref{tab:gnn_vs_settransformer}, the Set Transformer achieved competitive or superior region classification accuracy and point error compared to the GNN baseline, despite not relying on handcrafted graph structures or anatomical priors. 
In addition to accuracy, the Set Transformer demonstrated greater training stability and adaptability to new data distributions. 
These improvements are attributed to its attention-based modeling of raw circulation time sets, which preserves high-resolution temporal variability that is otherwise lost during feature summarization in the GNN pipeline.

Table~\ref{tab:gnn_vs_settransformer_augmentation} reveals that synthetic data augmentation did not further improve localization performance in the Set Transformer models, but it did slightly improve the region accuracy for GNN models. We argue that this is due to the fact that the Set Transformer attention heads work directly with raw data, and might extract more fine-grained patterns that are not shared among the synthetic data. On the other hand, GNN models work with aggregated statistical features, and therefore might benefit from additional data if the synthetic data distribution is generally similar.


\section{Conclusion}

This work investigates the potential of Set Transformer architectures and synthetic data generation techniques to enhance region-level localization within the bloodstream. Our motivation stems from the limitations of existing \ac{GNN}-based approaches~\cite{bartra2024graph}, which rely on handcrafted spatial priors and struggle to generalize to anatomical variability or limited data regimes. By treating nanodevice circulation time reports as unordered sets, we propose a formulation aligned with permutation-invariant architectures.

We extend the evaluation framework by exploring deep generative models, including \ac{CGAN}, \ac{WGAN}, \ac{WGAN-GP}, and \ac{CVAE}, to synthesize realistic circulation time samples. These models demonstrate strong distributional alignment with real data and were leveraged for data augmentation.

Our findings show that Set Transformer-based models can match or exceed \ac{GNN}-based baselines in terms of classification accuracy, with fewer architectural constraints and improved adaptability. However, limitations persist: the models still struggle to distinguish symmetric regions and may suffer from overfitting under sparse sampling conditions. Future work will explore hybrid models that combine the structural priors of \ac{GNN}s with the input flexibility of Set Transformers, and extend the evaluation to point-level localization accuracy under more realistic physiological scenarios.

\begin{table}[!t]
\centering
\caption{Performance of GNN and Set Transformer models without data augmentation}
\vspace{-2mm}
\small
\label{tab:gnn_vs_settransformer}
\begin{tabular}{@{}lccc@{}}
\toprule
Model & Accuracy~[\%] & Point error~[m] & Time~[s] \\
\midrule
\multicolumn{4}{c}{\textbf{GNN models}} \\
\midrule
GNN\_1 & \textbf{37.21} & \textbf{0.1249} & \textbf{262.50} \\
GNN\_2 & 35.79 & 0.1349 & 296.60 \\
GNN\_3 & 36.18 & 0.1354 & 292.25 \\
GNN\_4 & 35.40 & 0.1347 & 322.20 \\
GNN\_5 & 35.27 & 0.1329 & 321.25 \\
\midrule
\multicolumn{4}{c}{\textbf{Set Transformer models}} \\
\midrule
SetTransformer\_1 & \textbf{39.28} & 0.0993 & 2047.29 \\
SetTransformer\_2 & 36.43 & 0.1051 & 3263.50 \\
SetTransformer\_3 & 38.37 & 0.0994 & 2757.88 \\
SetTransformer\_4 & 38.37 & \textbf{0.0991} & \textbf{1754.72} \\
SetTransformer\_5 & 36.43 & 0.1051 & 2428.68 \\
\bottomrule
\end{tabular}
\vspace{-2mm}
\end{table}

\begin{table}[!t]
\centering
\caption{Performance of GNN and Set Transformer models with data augmentation}
\vspace{-2mm}
\small
\label{tab:gnn_vs_settransformer_augmentation}
\begin{tabular}{@{}lcc@{}}
\toprule
Augmentation type & Accuracy~[\%] & Point error~[m] \\
\midrule
\multicolumn{3}{c}{\textbf{GNN models}} \\
\midrule
None & 37.21 & \textbf{0.1249} \\
CGAN (Fixed) & 32.43 & 0.1577 \\
CVAE (Fixed) & \textbf{38.50} & 0.1339 \\
CGAN (Random) & 37.85 & 0.1354 \\
CVAE (Random) & 37.73 & 0.1337 \\
\midrule
\multicolumn{3}{c}{\textbf{Set Transformer models}} \\
\midrule
None & \textbf{39.28} & \textbf{0.0993} \\
CGAN (Fixed) & 29.97 & 0.1132 \\
CVAE (Fixed) & 32.04 & 0.1037 \\
CGAN (Random) & 33.72 & 0.1130 \\
CVAE (Random) & 34.11 & 0.1083 \\
\bottomrule
\end{tabular}
\end{table}

\section*{Acknowledgments}
This work was supported by the Generalitat de Catalunya's CERCA Programme.
Ethungshan Shitiri was supported by the EU Horizon Europe's Marie Skłodowska Curie programme (gr nº 10115485).

\renewcommand{\bibfont}{\footnotesize}
\printbibliography

\end{document}